\documentclass[russian,12pt]{article}
\usepackage[T1]{fontenc}
\usepackage[cp866]{inputenc}
\usepackage{babel}
\usepackage{psfig}
\makeatletter
\renewcommand\@biblabel[1]{}
\makeatother
\begin{document}
\oddsidemargin 15mm
\large
\baselineskip 0.76cm

УДК 550.38
\bigskip

\begin{center}
{\bf УСТАНОВЛЕНИЕ КОНВЕКЦИИ ВО ВРАЩАЮЩЕМСЯ СЛОЕ ВЯЗКОЙ ЖИДКОСТИ\\
С НАЛОЖЕННЫМ МАГНИТНЫМ ПОЛЕМ: ЗАВИСИМОСТЬ ОТ ЧИСЕЛ ПРАНДТЛЯ}

~

\copyright 2010~г. О.М.Подвигина

~

{\it Международный институт теории прогноза землетрясений\\
и математической геофизики РАН, г. Москва}
\end{center}

Изучено установление конвекции в горизонтальном слое проводящей несжимаемой
жидкости с жесткими диэлектрическими границами, подогреваемом снизу и
вращающемся относительно вертикальной оси, с наложенным вертикальным магнитным
полем в приближении Буссинеска. В зависимости от величин параметров задачи
(чисел Тейлора, Чандрасекара, кинематического и магнитного чисел Прандтля)
потеря устойчивости состояния покоя при увеличении числа Рэлея может
происходить с развитием монотонной или колебательной неустойчивости.
Конвективные валы, появляющиеся при монотонной неустойчивости, сами также
неустойчивы, если число Тейлора достаточно велико (имеет место так называемая
неустойчивость Кюпперса-Лорца). В данной работе исследовано, как критическое
значение числа Рэлея, тип неустойчивости состояния покоя и критическое значение
числа Тейлора для неустойчивости Кюпперса-Лорца зависят от кинематического и
магнитного чисел Прандтля. Рассмотрены числа Прандтля не превосходящие 1,
характерные для внешнего ядра Земли.

\bigskip\noindent
{\it Ключевые слова:} конвекция, числа Прандтля, магнитное поле,
монотонная и колебательная неустойчивости, неустойчивость\break
Кюпперса-Лорца

\bigskip
Согласно современным научным представлениям, источником магнитного поля Земли
является конвективное движение во внешнем ядре. Подтверждением этой гипотезы
служат, например, результаты серии работ Глатцмайера с соавторами (см.
[Glatzmaier, 1997; Christensen, 1999] и приведенные там ссылки), которые численно
решали систему уравнений, описывающую конвективные гидромагнитные явления.
В этих расчетах удалось воспроизвести дипольную в главном морфологию магнитного поля
Земли и его хаотические инверсии. Однако значения некоторых параметров,
использованных в этих работах, на несколько порядков отличаются от их значений
для ядра Земли, поэтому такое качественное соответствие
результатов расчетов с реальным геодинамо следует ``считать удивительным''
[Jones, 2000]. В частности, расчеты проводили для значения чисел Прандтля
порядка единицы, в то время как в расплавленном ядре кинематическое число
Прандтля имеет порядок $10^{-3}$, а магнитное -- $10^{-8}$ [Aurnou, 2001].
Отметим также, что разные авторы дают различные оценки этих величин для ядра
Земли, отличающиеся на несколько порядков (см. обсуждение этого вопроса в гл.~7
монографии [Merril, 1996]). Тем самым, представляет интерес исследовать, как
изменяется характер поведения конвективных течений при изменении значений чисел
Прандтля.

В настоящей статье задача о зависимости конвективных течений от кинематического и
магнитного чисел Прандтля рассмотрена в наиболее простой постановке. Мы
изучаем установление конвекции в горизонтальном слое проводящей жидкости,
подогреваемом снизу и вращающемся относительно вертикальной оси, с жесткими
диэлектрическими горизонтальными границами, поддерживаемыми при фиксированных
температурах, и с наложенным вертикальным магнитным полем. В безразмерной форме
система характеризуется числами Рэлея $R$ (относительная величина сил
плавучести), Прандтля $P$ (отношение кинематической вязкости к коэффициенту
тепловой диффузии), магнитного Прандтля $P_m$ (отношение кинематической
вязкости к коэффициенту магнитной диффузии), Тейлора $Ta$ (квадратный корень из
которого пропорционален скорости вращения) и Чандрасекара $Q$ (квадратный
корень из которого пропорционален интенсивности магнитного поля).

При малых числах Рэлея жидкость неподвижна. При увеличении числа Рэлея выше
некоторой критической величины возникает течение жидкости. При отсутствии и
вращения, и магнитного поля имеет место монотонная неустойчивость состояния
покоя жидкости [Chandrasekhar, 1961; Гершуни 1972; Гетлинг, 1991], возникающее
течение двумерно и имеет вид валов. При наличии магнитного поля, в зависимости
от величины отношения $P_m/P$, возможен качественное различный характер
возникающей неустойчивости [Подвигина, 2009; Podvigina, 2010]. Если $P_m<P$ (это
соотношение выполнено для величин параметров во внешнем земном ядре, и в данной
статье рассмотрен этот случай), при малых значениях $Ta$ при установлении
конвекции возникают устойчивые валы. При б\'ольших значениях числа Тейлора,
возникающие валы неустойчивы относительно возмущений, представляющих собой
такое же течение, повернутое на некоторый угол относительно вертикальной оси.
При отсутствии магнитного поля эту неустойчивость (ее называют неустойчивостью
Кюпперса-Лорца) исследовали K\"uppers and Lortz [1969] и Clune and Knobloch [1993].
При дальнейшем увеличении скорости вращения монотонная неустойчивость состояния
покоя сменяется колебательной.

В статье рассмотрены числа Прандтля на интервалах \hbox{$10^{-3}\!\le\! P\le1$}
и $10^{-8}\le P_m\le 1$. Тем самым, минимальные значения приблизительно отвечают
значениям параметров во внешнем ядре Земли, а максимальные -- используемым при
численном моделировании происходящих там процессов. Наши расчеты проведены для
чисел Тейлора и Чандрасекара, не превышающих $10^{10}$ и $2\cdot10^4$,
соответственно, а в земном ядре этих величины $\sim10^{24}$ и $\sim10^{10}$
[Aurnou, 2001]. Однако, как отмечал еще Chandrasekhar [1961], при $Ta\to\infty$
и $Q\to\infty$ поведение системы носит асимптотический характер, и потому
экстремально большие значения этих параметров не представляют интереса.
Численно исследована зависимость критического числа Рэлея, типа неустойчивости
состояния покоя, критических значений числа Тейлора и угла неустойчивости
Кюпперса-Лорца от чисел Прандтля. Для большого числа рассмотренных случаев
изменение одного из чисел Прандтля на порядок или несколько порядков приводит к
незначительным изменениям рассматриваемых критических значений. Тем самым,
система несущественно зависит от чисел Прандтля, что, возможно, объясняет
качественное сходство магнитного поля, полученного в упомянутых численных
экспериментах, с магнитным полем Земли. (Отметим принципиальные сложности,
связанные с тем, что, с одной стороны, проводить расчеты с достаточным
разрешением при значениях параметров, соответствующих условиям во внешнем ядре
Земли, существующий уровень развития вычислительной техники не позволяет,
а с другой, достаточно детальное сравнение результатов расчетов с геомагнитным
полем невозможно из-за недостаточности информации о структуре магнитного поля.)
Наши результаты показывают, что ошибки в оценках этих параметров могут быть
несущественными.

\bigskip\noindent
{\bf 1. Уравнения и параметры}

\bigskip
Эволюция конвективного течения и магнитного поля описывается уравнениями
Навье-Стокса, теплопроводности и магнитной индукции. В безразмерном виде они
имеют вид
\bigskip
$$
{\partial{\bf v}\over\partial t}={\bf v}\times(\nabla\times{\bf v})
+P\Delta{\bf v}+PR\theta{\bf e}_z-\nabla p$$
$$
+\,PTa^{1/2}{\bf v}\times{\bf e}_z
+P^2P_m^{-1}Q(\nabla\times{\bf B})\times{\bf B}
\eqno(1)
$$
$$
{\partial\theta\over\partial t}=-({\bf v}\cdot\nabla)\theta+v_z+\Delta\theta
\eqno(2)
$$
$${\partial{\bf B}\over\partial t}=\nabla\times({\bf v}\times{\bf B})
+PP_m^{-1}\Delta{\bf B}
\eqno(3)
$$

Поле скорости и магнитное поле бездивергентны:
$$\nabla\cdot{\bf v}=0,\qquad\nabla\cdot{\bf b}=0
\eqno(4)
$$

\bigskip
В этих уравнениях $\bf v$ -- скорость потока, $\theta$ -- разность между
температурой и ее линейным профилем, устанавливающемся в слое в состоянии
покоя, $\bf b$ -- разность между магнитным полем и его постоянной вертикальной
составляющей, ${\bf B}={\bf e}_z+{\bf b}$, ${\bf e}_z$ -- единичный вектор
вдоль вертикальной оси $z$. На горизонтальных границах выполнено условие
прилипания и фиксированы температуры:
\bigskip
$$
v_x=v_y=v_z=0,\qquad\theta=0\qquad(z=\pm{1\over 2})
\eqno(5)
$$

Считаем, что внешняя среда за пределами слоя -- изолятор, т.е. магнитное поле
на границе равно градиенту гармонической функции, определенной в
соответствующем полупространстве [Chandrasekhar, 1961]:
$$
{\bf b}(z)=\nabla\phi(z)\ (z=\pm{1\over 2})
\eqno(6)
$$
$$\Delta\phi=0\ (|z|>{1\over 2}),\qquad\phi\to0\ (|z|\to\infty)$$

Безразмерные параметры определяются следующими соотношениями:
$$
P={\nu\over\kappa},\ P_m={\nu\over\mu},\
R={\alpha g d^3\over\nu\kappa}\delta T,\
Ta=({2\Omega d^2\over\nu})^2,\ Q={\sigma B_0^2d^2\over\rho\nu}
$$
Здесь $\nu$ -- кинематическая вязкость, $\kappa$ -- коэффициент термической
диффузии, $\mu$ -- коэффициент магнитной диффузии, $\alpha$ -- коэффициент
термического расширения, $g$ -- ускорение свободного падения, $d$ -- толщина
слоя, $\delta T$ -- разность температур между верхней и нижней границами,
$\Omega$ -- скорость вращения, $\sigma$ -- коэффициент электропроводности,
$B_0$ -- интенсивность внешнего магнитного поля, $\rho$ -- плотность жидкости.

\bigskip\noindent
{\bf 2. Устойчивость состояния покоя}

\bigskip
Устойчивость стационарного состояния $({\bf v}=0,\ {\bf b}=0,\ \theta=0)$
(жидкость неподвижна) определяется собственными значениями оператора
линеаризации (1)--(3). Стационарное состояние становится\break неустойчивым, когда
собственное значение этого оператора пересекает мнимую ось.

Вычисление критических значений следует методу, впервые использованному в
[Pellew, 1940] для расчета критических значений Рэлея состояния равновесия
жидкости в горизонтальном слое без вращения и магнитного поля и примененного
в [Clune, 1993; Weeks, 2002; Zhang, 2004] для конвективной системы с вращением
и/или магнитным полем. Для фиксированных параметров ($P$, $P_m$, $Ta$ и $Q$)
критическую моду ищем в виде конечной суммы произведений тригонометрических
функций и экспонент с неизвестными коэффициентами. Подставляя эту сумму
в линеаризованные уравнения конвекции, теплопроводности и магнитной индукции
и в граничные условия, получим систему уравнений (она выведена в [Podvigina,
2010]), из которой находим критическое число Рэлея $R_c(k)$ и частоту (для
колебательной неустойчивости), зависящие от $k$ -- горизонтального волнового
числа критической моды. Из-за сложности этих уравнений, систему решаем
численно. Определяя минимум $R_c(k)$ по $k$ и сравнивая значения для разных
типов неустойчивости, находим тип критической моды, частота $\omega_c$ (для
колебательной неустойчивости) и критическое значение $R_{cr}$ для данных $P$,
$P_m$, $T$ и $Q$.

Зависимости $R_c^s(k)$ и $R_c^o(k)$ (критические значения чисел Рэлея для
монотонной и колебательной неустойчивостей, соответственно) от $k$ при
$Ta=10^6$, $Q=100$ и нескольких значениях $P$ и $P_m$ показаны на рис.~1. При
данном $k$, для монотонной неустойчивости $R_c^s(k)$ зависит только от $Ta$
и $Q$, для колебательной $R_c^o(k)$ также зависит от чисел Прандтля.
Колебательная неустойчивость возможна при достаточно малых $P$, $R_c^o(k)$
растет с ростом числа Прандтля и убывает с ростом магнитного числа Прандтля.
Для малых $P$ и $P_m$ эта зависимость оказывается весьма слабой. Например,
на рис.~1а графики, соответствующие $P_m=10^{-3}$ и $P_m=10^{-8}$, зрительно
неразличимы. На рис.~1б графики, отвечающие $P=10^{-2}$ и $P_m=10^{-3}$,
также отличаются несущественно. Слабость зависимости характерна для всех
рассмотренных значениях $Ta$ и $Q$ (а именно, $Ta=2.5\cdot10^5$, 1000, 2000
и $Q=200$, 500, 2000, 5000).

Зависимость $R_{cr}$ от $P_m$ для нескольких значений $Q$ показана на рис.~2а,
при этих значениях параметров имеет место колебательная неустойчивость состояния
покоя. В диапазоне $10^{-6}\le P_m\le10^{-2}$ (т.е. при $P_m\le P$) изменения
$R_c^o$ незначительны, эта величина меняется существенно только при $P_m$
порядка 0.1\,. При небольших $P$, $R_{cr}$ также зависит от $P$ слабо (см. рис.~2б).
Горизонтальные отрезки графиков (правая часть рассмотренного интервала $P$)
отвечают монотонной неустойчивости (т.е. изображают $R_{cr}^s$), а их продолжения
слева -- колебательной (т.е. левые фрагменты графиков изображают $R_{cr}^o$).
Известно [Chandrasekhar, 1961], что $R_{cr}^s$ убывает с ростом $Q$ при больших $Ta$
и малых $Q$; для колебательной неустойчивости имеет место обратная зависимость
от $Q$ (см. рис.~2б).

Зависимость $R_{cr}$ и $\omega_{cr}$ от $Ta$ показана на рис.~3 для нескольких
значений $Q$ и $P_m$. Монотонная неустойчивость сменяется колебательной
при достаточно больших числах Тейлора. Отметим, что графики, отвечающие
$P_m$, отличающимся на шесть порядков, -- $10^{-8}$ и $5\cdot10^{-3}$ --
зрительно неразличимы. При больших $Ta$ зависимости $R_{cr}$ и $\omega_{cr}$
выходят на асимптотический режим поведения.

\bigskip\noindent
{\bf 3. Неустойчивость Кюпперса-Лорца}

\bigskip
Для вычисления критических значений $Ta$ и угла между осями возмущаемых валов
и возмущением, $\alpha$, применен метод [Podvigina, 2010], основанный
на использовании амплитудных уравнений, аналогичный примененному в [Clune, 1993]
для исследования устойчивости валов во вращающемся слое. В горизонтальной
плоскости предполагается, что векторы периодов системы возмущенных валов
и возмущения образуют ромбическую ячейку периодичности с углом $\alpha$.
Используя методы теории динамических систем (а именно, ограничение на
центральное многообразие), можно вычислить коэффициенты амплитудных уравнений.
Валы устойчивы относительно возмущений, отвечающих некоторому $\alpha$, если
для этого $\alpha$ коэффициенты амплитудных уравнений удовлетворяют определенным
неравенствам (см.~[Podvigina, 2010]). Если эти неравенства выполнены для всех
$\alpha$, $0\le\alpha\le\pi/2$, то конвективные валы устойчивы относительно
возмущений рассматриваемого типа.

Расчеты проведены для $Q=5$,50, 500 и 5000 с шагом $\pi/500$ по $\alpha$.
Типичная зависимость $Ta_c$ и $\alpha_c$ от
числа Прандтля показана на рис.~4. При достаточно малых $P_m$ эта зависимость
такая же, как и во вращающемся слое без внешнего магнитного поля:
эти величины строго положительны и монотонно растут с ростом $P$. При б\'ольших
$P_m$ поведение иное, неустойчивость может иметь место уже при $Ta=0$.
Однако для $P_m<P$ (напомним, что для геофизических приложений представляет
интересен именно этот случай) зависимость от $P_m$ незначительна.

\bigskip\noindent
{\bf Заключение}

\bigskip
Результаты приведенных выше расчетов показывают, что в рассмотренных диапазонах значений
параметров ($0.001\le P\le1$, $10^{-8}\le P_m\le1$, $P_m<P$) степень
зависимости критических параметров (числа Рэлея и частоты при колебательной
неустойчивости состояния покоя, числа Тейлора и угла для неустойчивости
Кюпперса-Лорца) от магнитного числа Прандтля крайне мала.
Как правило, изменение $P_m$ на 5-7 порядков влечет изменение
критических значений менее чем на 1\%. Для достаточно малых $P$ зависимость
от этого параметра незначительна: так, при уменьшении $P$ от 0.01 до 0.001
критические значения меняются несущественно.

Эти результаты указывают на существование асимптотических режимов при
$P\to0$ и $P_m\to0$, малость коэффициентов асимптотических
рядов и ненулевой радиус их сходимости. (Построение асимптотических
разложений выходит за рамки настоящей статьи.)

Отметим, что уравнения, геометрия, величины чисел Рэлея, Тейлора и Чандрасекара
рассмотренной задачи отличны от условий Земли, и, таким образом, изученные
в данной статье магнитогидродинамические конвективные режимы отличны от
реализующихся в геофизических конвективных системах. Однако наблюдаемые
зависимости от чисел Прандтля имеют достаточно общий характер, поскольку
выполняются во всех рассмотренных нами случаях; эта универсальность дает
основания полагать, что нечувствительность магнитогидродинамических конвективных
режимов может сохраниться и при величинах параметров и режимах, характеризующих
условия Земли.

Работа поддержана грантами ANR-07-BLAN-0235 OTARIE Агентства национальных
исследований Франции и 07-01-92217-НЦНИЛ\_а РФФИ.

\pagebreak

\pagebreak

\begin{figure}
\vspace*{-34mm}
\centerline{
\psfig{file=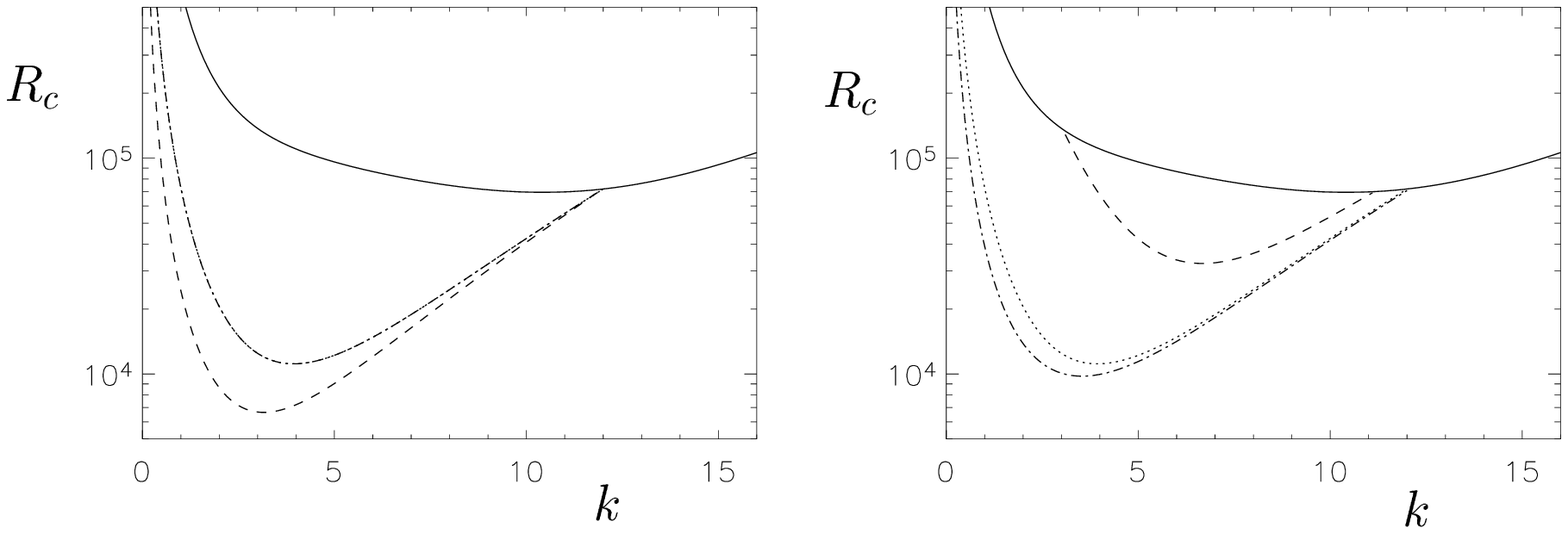,width=18cm}
}
\vspace*{-164mm}
\centerline{
(а)\hspace{78mm}(б)
}

\caption{
Зависимость критического числа Рэлея $R_c$ от горизонтального волнового
числа нейтральной моды $k$ при $Ta=10^6$, $Q=100$: (а) $P=0.01$, $P_m=0.1$
(штриховая линия), $P_m=10^{-3}$ (пунктир), $P_m=10^{-8}$ (штрих-пунктир);
(б) $P_m=10^{-4}$, $P=0.1$ (штриховая линия), $P=0.01$ (пунктир), $P=10^{-3}$
(штрих-пунктир). Сплошная линия -- стационарная мода, пунктир, штриховая линия
и штрих-пунктир -- осциллирующая.
}
\label{fig51}
\end{figure}

\begin{figure}
\vspace*{-34mm}
\centerline{
\psfig{file=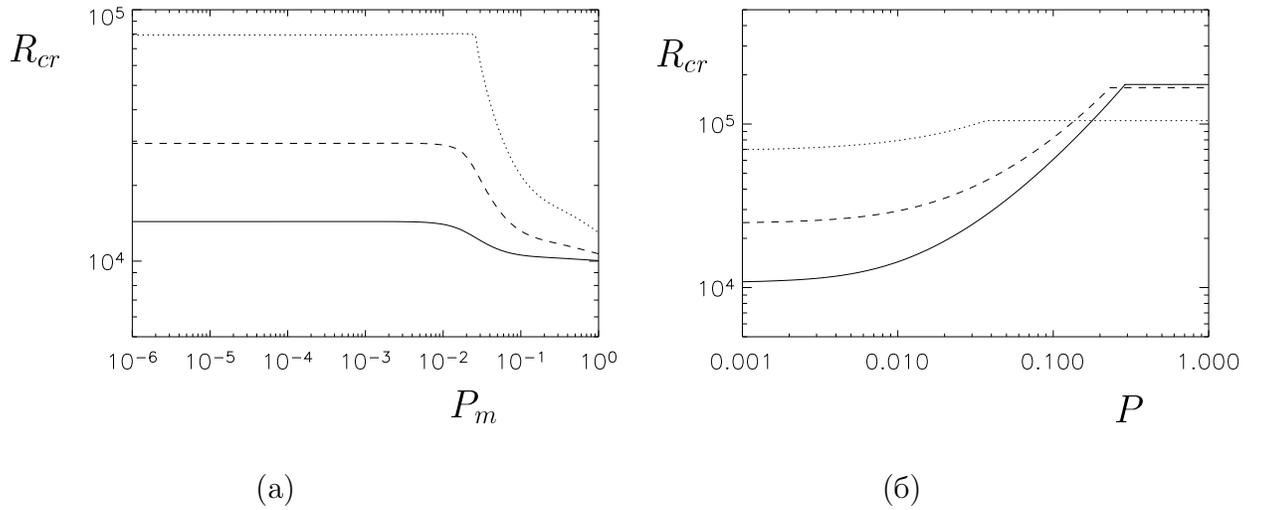,width=18cm}
}
\vspace*{-164mm}
\centerline{
(а)\hspace{78mm}(б)
}

\caption{
Зависимость критического числа Рэлея $R_{cr}$ от магнитного (а)
и кинематического (б) чисел Прандтля при $Ta=4\cdot10^6$ и $Q=100$ (сплошная
линия), 500 (штриховая), 2000 (пунктир): (а) $P=0.01$, (б) $P_m=10^{-4}$.
}
\label{fig52}
\end{figure}

\begin{figure}
\vspace*{-34mm}
\centerline{
\psfig{file=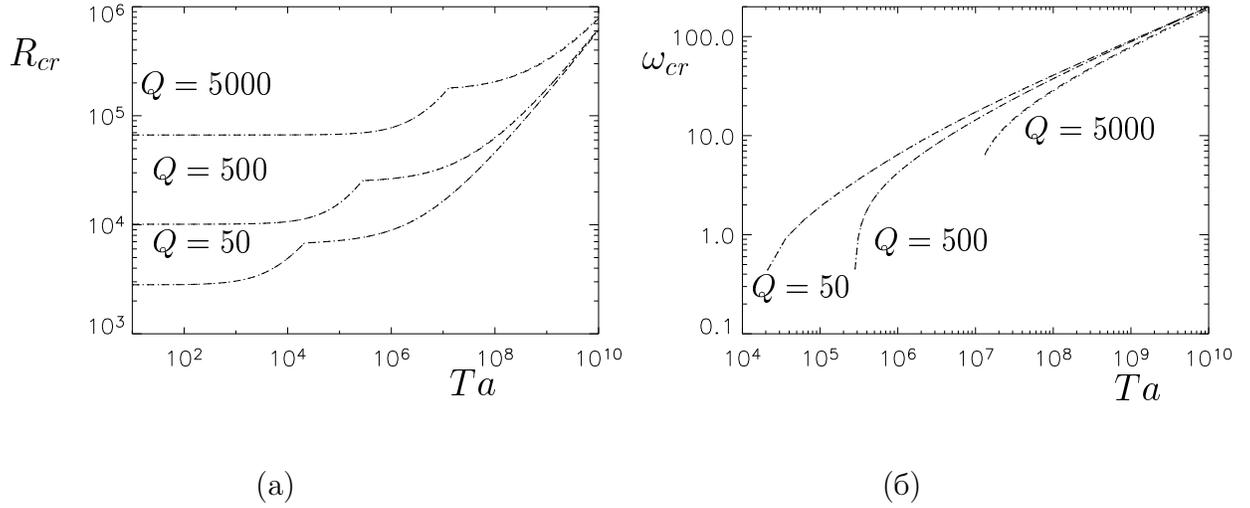,width=18cm}
}
\vspace*{-164mm}
\centerline{
(а)\hspace{78mm}(б)
}

\caption{
Зависимость критического числа Рэлея $R_{cr}$ (а) и частоты
$\omega_{cr}$ (б) от числа Тейлора для $P=0.01$ и трех величин $Q$ при
$P_m=5\cdot10^{-8}$ (пунктир) и $P_m=10^{-8}$ (штриховая линия).
}
\label{fig53}
\end{figure}

\begin{figure}
\vspace*{-34mm}
\centerline{
\psfig{file=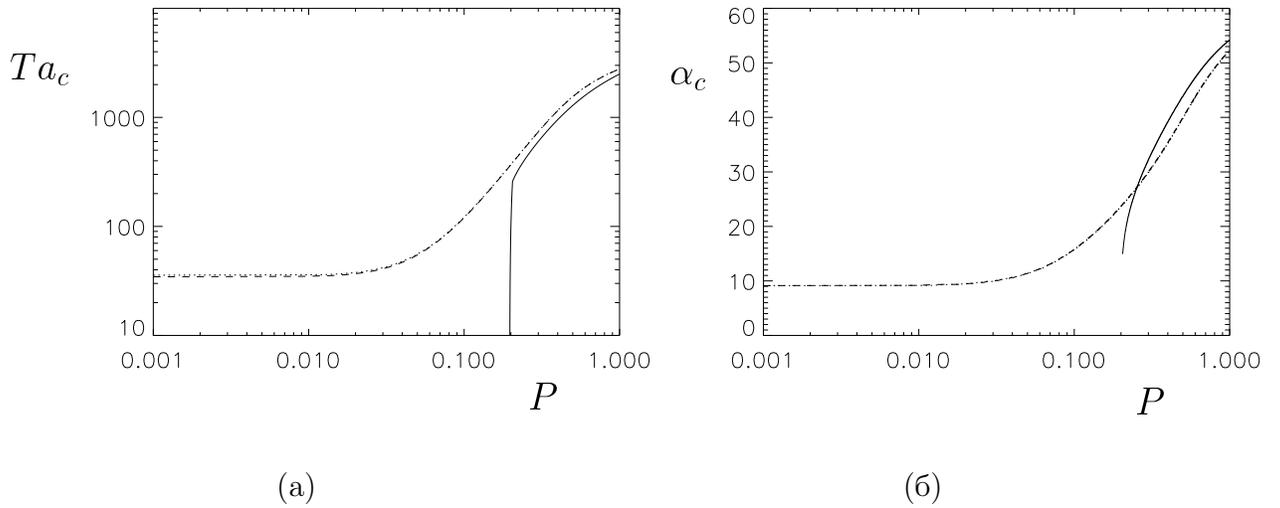,width=18cm}
}
\vspace*{-164mm}
\centerline{
(а)\hspace{78mm}(б)
}

\caption{
Зависимость критического числа Тейлора $Ta_c$ (а) и угла
$\alpha_c$ (б) от числа Прандтля для $Q=500$, $P_m=0.1$ (сплошная линия),
$P_m=10^{-2}$ (пунктир) и $P_m=10^{-8}$ (штриховая линия).
}
\label{fig54}
\end{figure}
\end{document}